\documentclass[conference]{IEEEtran}

\usepackage{subfigure}
\usepackage{graphicx}
\usepackage{courier}
\usepackage{balance}
\def\BibTeX{{\rm B\kern-.05em{\sc i\kern-.025em b}\kern-.08em T\kern-.1667em\lower.7ex\hbox{E}\kern-.125emX}}

\ifCLASSINFOpdf
\else
\fi

\IEEEoverridecommandlockouts 
\begin{document}

\title{CSAI: Open-Source Cellular Radio Access Network Security Analysis Instrument}

\author{\IEEEauthorblockN{Thomas Byrd and Vuk Marojevic}
\IEEEauthorblockA{Dept. of Electrical and Computer Engineering\\
Mississippi State University\\
Mississippi State, MS\\
\{tkb140, vuk.marojevic\}@msstate.edu}
\and
\IEEEauthorblockN{Roger Piqueras Jover\IEEEauthorrefmark{1} \thanks{\IEEEauthorrefmark{1}Author did not contribute to source code. Code published to accompany this paper was written by Mr. Thomas Byrd and Dr. Vuk Marojevic.}}
\IEEEauthorblockA{Bloomberg LP\\
New York, NY\\
rpiquerasjov@bloomberg.net}
}

% make the title area
\maketitle

% As a general rule, do not put math, special symbols or citations
% in the abstract
\begin{abstract}
This paper presents our methodology and toolbox that allows analyzing the radio access network security of laboratory and commercial 4G and future 5G cellular networks. We leverage a free open-source software suite that implements the LTE UE and eNB enabling real-time signaling using software radio peripherals.  We  modify the UE software processing stack to act as an LTE packet collection and examination tool. This is possible because of the openness of the 3GPP specifications. Hence, we are able to receive and decode LTE downlink messages for the purpose of analyzing potential security problems of the standard. This paper shows how to rapidly prototype LTE tools and build a software-defined radio access network (RAN) analysis instrument for research and education. Using CSAI, the Cellular RAN Security Analysis Instrument, a researcher can analyze broadcast and paging messages of cellular networks. CSAI is also able to test networks to aid in the identification of vulnerabilities and verify functionality post-remediation. Additionally, we found that it can crash an eNB which motivates equivalent analyses of commercial network equipment and its robustness against denial of service attacks. 
\end{abstract}

%keywords

\begin{IEEEkeywords}
open-source LTE, SDR, paging, radio access network signaling, analysis, wireless security
\end{IEEEkeywords}

\IEEEpeerreviewmaketitle

\section{Introduction}

The Long Term Evolution (LTE) is a cellular communications standard developed by the 3rd Generation Partnership Project (3GPP). LTE was finalized in 3GPP Release 8 in December 2008, and LTE-Advanced followed in 3GPP Release 10. Only recently has there been significant enough open source software development efforts for producing stable implementations of the LTE and LTE-A specifications to allow for rapid prototyping and testing of 4G networks by the broader research community.

Next generation 5G networks promise a huge leap from 4G. The reality however is that the initial 5G releases leverage LTE networks in many regards: New Radio (NR) initially implements a similar radio access network (RAN) and hooks to the LTE evolved packet core (EPC). 5G generally allows more flexible waveform and protocol configurations, transmission in sub 6 GHz and millimeter wave bands, and higher bandwidths than LTE. The signaling will initially be OFDM, for example, where the 5G signaling frame will carry data and control information. Starting in 3GPP Release 15, 5G frames, channels, and signals are specified in the standards specifications.

There is a huge need for research and development tools that enable cellular signaling analysis for a multitude of purposes. It can help understand the limitations of current implementations and guide the evolution of the standard. They can also be effectively used for education and training. Security is another important aspect where RAN signaling analysis is needed. It has been shown that the LTE control signaling suffers from targeted interference that an adversary can exploit, easily and cheaply \cite{lichtman2016lte}. We therefore propose a flexible signal analysis tool for analyzing commercial and experimental cellular communication systems, assisting in the detection of potential vulnerabilities, and evaluating corrective measures which will pave the path to secure wireless networks.

This paper leverages open-source software implementations of LTE and develops a free open-source cellular RAN security analysis instrument, CSAI. CSAI is lightweight and can process data in real time. It interfaces with common software radio front ends, such as Ettus Research USRPs, and can capture LTE control messages and be easily extended to capture 5G NR signals. It can emulate an eNodeB (eNB) or user equipment (UE) and implement specific processes to test the behavior of the UE or eNB. It also allows testing larger RANs which involve multiple UEs or multiple eNBs. For example, in commercial networks that have dozens of UEs, or more, that rotate between serving cells, this tool will be able to monitor paging traffic in a particular cell and identify new UEs as they are paged for the purpose of signaling analysis.

It is very important to be able to analyze protocol edge cases and understand their implications in terms of RAN security. Not only can it be used for analyzing the standard specifications of a modern cellular standard, but this tool can also test vendor specific implementations. Additionally, it is a benchmarking tool for stress testing 4G and 5G networks and can be adapted to fit different use cases. For instance, if a vendor needs an automated tool to determine the limits of their Radio Resource Control (RRC) buffers, this instrument will be able to facilitate that. 

The remainder of this paper is organized as follows. Section II briefly outlines other work in the area of capturing LTE  messages and performing LTE security analyses. Section III describes the important LTE signaling over the RAN. This allows for better comprehension of Section IV, which introduces our software instrument for analyzing broadcast and paging messages. Section V discusses experiments and data collected from commercial networks. Section VI focuses on the security implications of our initial results, and Section VII concludes the paper.

\section{Related Work}
Security research of cellular communications standards has a long history and helped evolve systems to the current 4G and emerging 5G networks \cite{jover2019security}. The insecure 2G systems are still used today and whenever 4G or 3G coverage is not available, handsets look for 2G networks. 4G systems introduce network and user authentication, where a user can authenticate the network it connects to. However, certain 4G security vulnerabilities were identified that 5G networks intend to fix.
	
With the emergence of software radios, increasing processing power of general-purpose computers, and software implementations of cellular standards, experimental LTE security research took off \cite{rao2017lte}. Researchers dissected the entire LTE signaling frame looking for vulnerabilities of the system when specific subsystems are interfered with. Two types of attacks were examined, physical control channel jamming and spoofing, and mitigation mechanisms were proposed in \cite{marojevic2017performance} and \cite{labib2017enhancing}. Other research groups have more recently tested LTE’s higher layer signaling protocols and published their findings in open literature \cite{hussain2018lteinspector}.	

While there exist many commercial tools that perform LTE traffic capture and decoding,  to our knowledge, there is no open source software that will accomplish this.  Papers that have been published regarding LTE security require the use of commercial LTE capture tools or the development of custom tools as observed in \cite{shaik2015practical} and \cite{jover2016lte}. The relevance of this subject is apparent from the availability of professional test instruments, offered by various hardware and software companies. But their high cost limits their widespread use in research and education.  Our goal is to provide a framework for making cellular RAN signaling analysis accessible to all, enable wireless security research, increase the transparency and visibility of RAN operations, and allow easy adoption by industry and standardization bodies.

\section{Background}

This section provides the necessary background on how LTE UEs register to the network and get notified by the network of incoming messages or calls. When a UE powers on, it first needs to receive and decode the Primary and Secondary Synchronization Signals (PSS/SSS) \cite{3gpp.36.211}.  Together, these two signals allow the UE to synchronize on a slot and frame level basis, respectively, as well as correct for frequency and phase offsets between the eNB and UE oscillators Now that the UE is synchronized with the eNB, it needs to know more information before it can initiate an attach request.  It needs to decode the the Master and System Information Blocks (MIB/SIBs). These blocks are transmitted in the clear by the eNB on a regular basis to ensure that UEs have the necessary information needed to attach. This is the initial cell search that each UE performs when turned on or when returning out of coverage and is part of the information that our tool can capture and analyze. 

\begin{figure}[htbp]
\centerline{\includegraphics[width=\columnwidth]{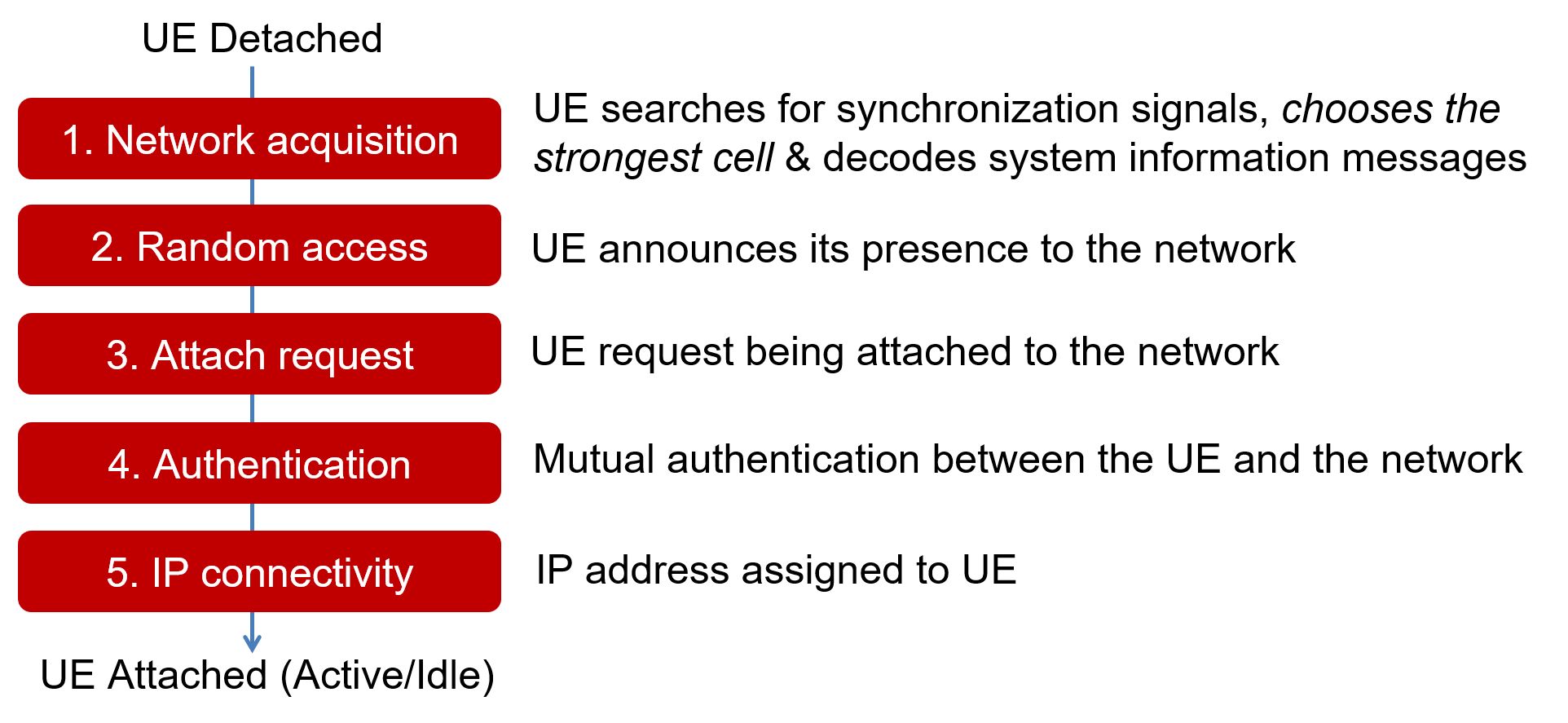}}
\caption{Life cycle of an LTE UE.}
\label{fig1}
\end{figure}

Once a UE knows the network configuration details that are provided in the MIB/SIBs, it can then use its Random Access Radio Network Temporary Identity (RA-RNTI) to initiate a RRC connection with an eNB. After a UE has established an RRC connection, the UE will communicate to the EPC through the eNB over the Non-Access Stratum (NAS) protocol layer. Fig. 1 illustrates this attachment process.

In order to identify itself with the network, UEs utilize the International Mobile Subscriber Identity (IMSI). This secret identifier can be leveraged in a number of privacy-invading attacks \cite{engel2008locating} and, as such, should always be kept private. However, the UE will authenticate with the EPC’s Home Subscriber Server (HSS) transmitting its IMSI in the clear if the UE has no history with the network. 

Once all of the NAS and RRC connections are established, the UE will enter an Idle state and deactivate the radio link between itself and the eNB. If the EPC needs to deliver a message to an idle UE, it is the job of the eNB to ‘wake-up’ the idle device and re-establish a physical connection \cite{linkedin}. This is done by sending out a paging message to all UEs in the operational area of the eNB. These paging messages use a specific Paging RNTI (P-RNTI) \cite{3gpp.36.321} to indicate the broadcast nature of paging and UEs are required to respond if their IMSI or SAE Temporary Mobile Subscriber Identity (S-TMSI) is being paged. The S-TMSI is a combination of MME Code and the Mobile TMSI (m-TMSI), herein both are simply referred to as the TMSI. Our tool is able to capture and decode the SIBs and Paging messages for specified eNBs which enables RAN Security analysis.

\section{CSAI: Cellular RAN Security Analysis Instrument}
There are various open source software applications that implement different parts of the LTE specifications. We chose srsLTE \cite{gomez2016srslte} for its simplicity and applicability toward capturing and decoding broadcast and paging messages.
	
srsLTE specifically implements 3GPP Release 8 with certain components of Release 9 integrated into its software. It is a licensed under the GNU Affero General Public License for free use for non-commercial purposes, such as research and education. The srsLTE software suite is compatible with software defined radio (SDR) hardware to build LTE radio access networks. As the names suggest, srsUE implements the LTE UE and srsENB the LTE eNB. To accompany these, Software Radio Systems (SRS) has also published srsEPC which provides an Evolved Packet Core (EPC) that is needed for a fully working  LTE network with one or several eNBs serving one or several UEs.  At the time of writing, we are using the most current version of srsLTE, version 18.12.0 based on commit 3cc4ca85 from the master branch \cite{srsLTE2019}.

There are two primary ways that a message capture program can be implemented using the srsLTE software suite. The first method has been used in many different research papers and uses one of srsLTE’s example binaries which requires little modification but significant external processing to generate useful data.  The second method requires more modification to srsUE, but automatically decodes data and presents it in an easily readable format.

\subsection{Method 1}
Method 1 involves using one of the example binaries. The one we are focusing on is \texttt{pdsch\_ue} which is one of many example programs provided for LTE network testing. Its companion program is \texttt{pdsch\_enodeb} which can generate PSS/SSS and MIB/SIBs. These blocks can be transmitted over a physical radio frequency (RF) interface or  written to a file.  The UE application is able to decode the synchronization signals as well as the information blocks.  It also has the capability to listen for a specific RNTI and only decode blocks addressed to that RNTI.  As defined in the Medium Access Control (MAC) protocol specifications [12], P-RNTIs have the fixed value 0xFFFE, and we can instruct \texttt{pdsch\_ue} to only listen for this specific paging channel RNTI.  The only modification we need to make to \texttt{pdsch\_ue} is one to get the raw information blocks and paging messages from the program. 

In order to achieve this, an additional line of code is needed after the application decodes the PDSCH to print the frame received from an RF interface or input file.  For the data to be written to a file, the necessary configuration code must be added before the program enters the main loop that continually receives and decodes PDSCH frames.  

The main problem with this approach is that additional work is required to decode the output data using an ASN.1 message decoder. An example output is shown in Figure 2. While there exist online tutorials and LTE message decoders, they either add unnecessary complication to the workflow or are inadequate for the amount of processing required for real time operations.  While this is a valid method to capture paging traffic, we did not want to hinder large-scale analysis and choose Method 2 for our tool.

\begin{figure}[htbp]
\centerline{\includegraphics[width=\columnwidth]{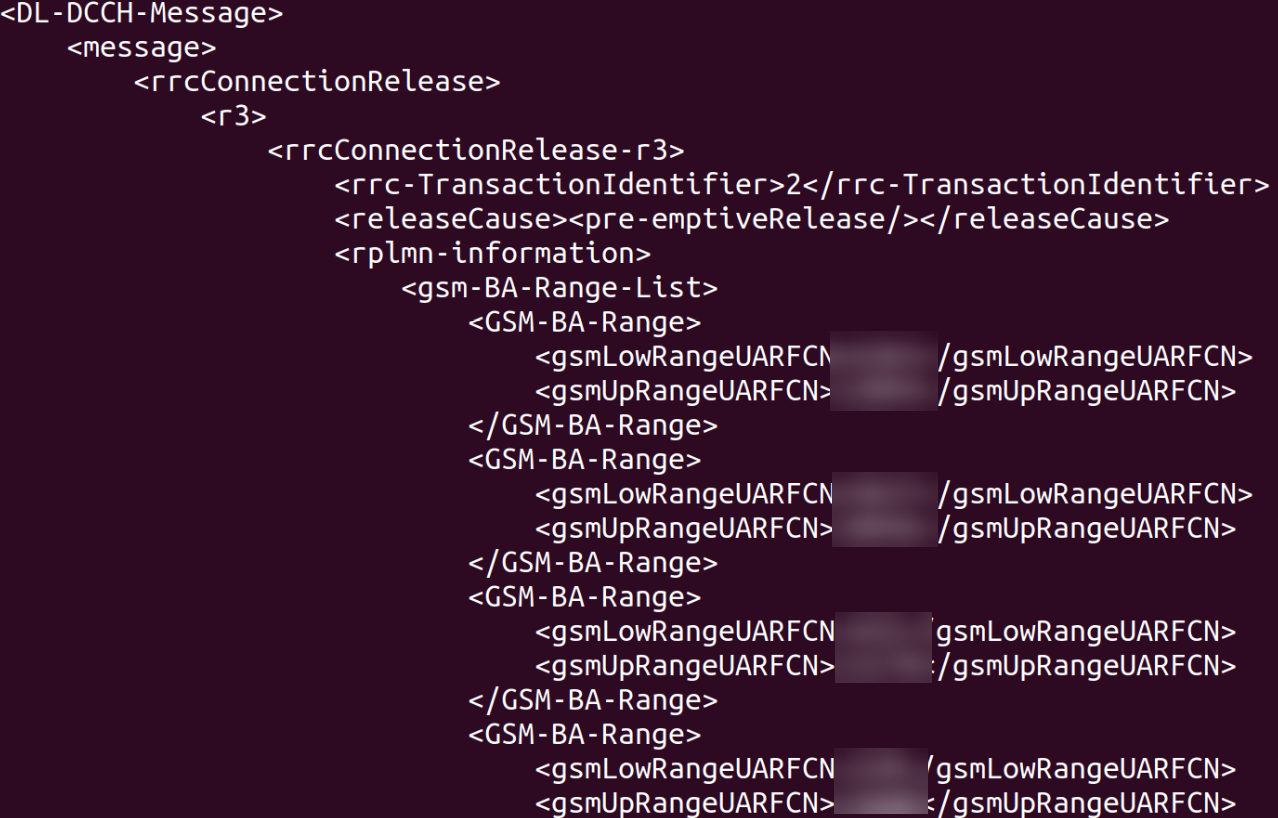}}
\caption{Output from ASN.1 Decoding}
\label{fig2}
\end{figure}

\subsection{Method 2}
The second method involves directly modifying the srsUE source to achieve the desired capabilities. Specifically, we modify the code that implements the RRC protocol. The RRC protocol is primarily responsible for connection establishment and release as well as handling paging messages.  In the \texttt{connection\_request} function, \texttt{send\_con\_request} is called which is responsible for sending the RRC connection request message to the lower layers that is transmitted to the eNB.  If we comment this function and replace it with a call to \texttt{rrc\_conection\_release}, we instruct the UE to remain disconnected and not communicate with an eNB.  This alone allows capturing the SIBs transmitted by the eNB, but is not enough to capture paging messages as well.  

In order to capture paging messages, we add an additional line after the connection release call to update the RRC state to reflect a successful connection. The other layers of srsUE will now look for paging messages and they are automatically captured and logged if configured to. 

Finally, the capture flag must be enabled. This is achieved by editing the \texttt{ue.conf} file to enable PCAP logging and set debug level logging for the MAC and RRC layers.  The log file will display all decoded SIBs and paging messages, but they are also available in the generated pcap file. We can view the capture in Wireshark by making an entry in the DLT\_USER encapsulation table.  The required DLT settings are listed in configuration files. After making all necessary modifications and rebuilding the software, outputs as shown in Figures 3 and 4 can be obtained for analysis. The next section describes some of the statistics we derived from using CSAI. 
\begin{figure}[htbp]
\centerline{\includegraphics[width=\columnwidth]{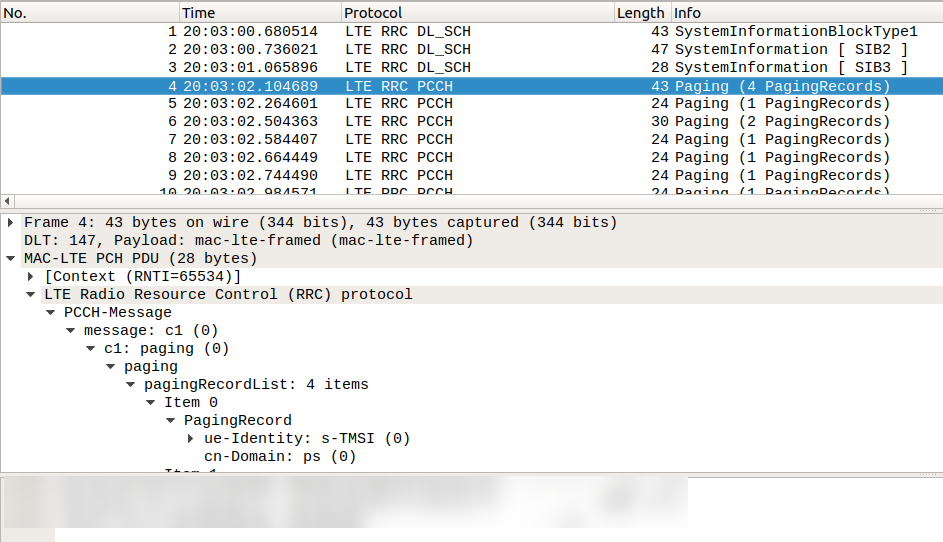}}
\caption{Wireshark analysis of paging messages.}
\label{fig3}
\end{figure}

\begin{figure}[htbp]
\centerline{\includegraphics[width=\columnwidth]{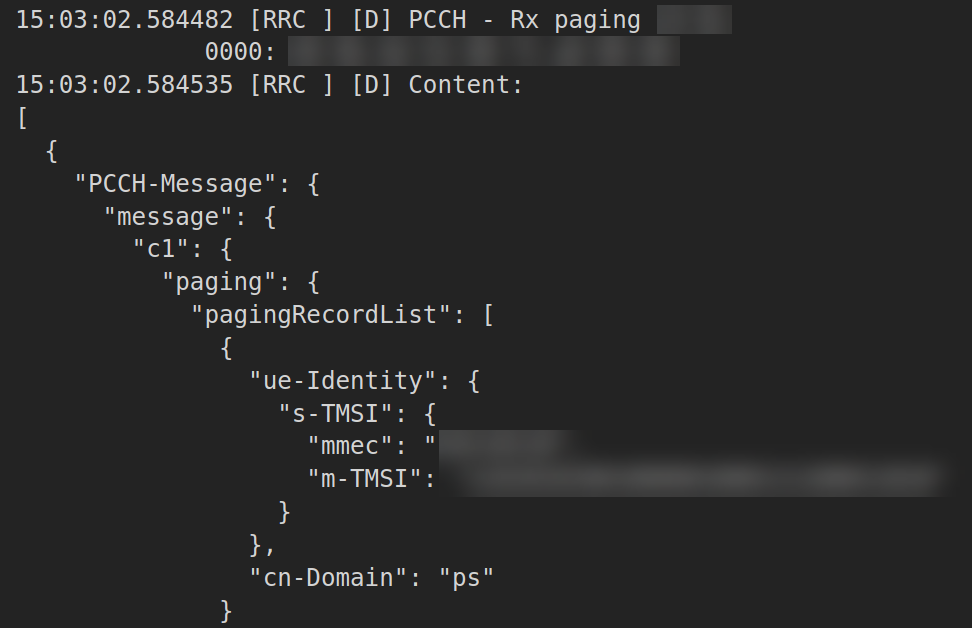}}
\caption{Log output of paging message.}
\label{fig4}
\end{figure}

\section{Experimentation}
In order to benchmark CSAI, we capture commercial network traffic and provide masked statistics to show the effectiveness of the tool. Two USRP B210s with the modifications detailed in Section IV were used to capture SIBs and paging traffic on an Ubuntu 18.04.02 machine. Similarly to \cite{hussain2019privacy} and \cite{chlosta2019lte}, only SIB and paging messages were acquired; careful consideration was taken to ensure that no user data was captured or retained despite being encrypted. 

\subsection{Short-Term Persistence}

Table I shows the data that we obtained from three networks. We measured the amount of total paging traffic over a six-hour time frame of two network operators the first day, and repeated the same capture for the third operator on the following day.  By inspecting the paging information to see whether IMSIs were used to page users, we found that all pages used S-TMSIs to identify a UE as opposed to revealing the IMSI, which is an encouraging result. 

The last row in Table I shows the longest observed TMSI in minutes which matches the length of the experiment.  In all three cases, TMSIs were observed throughout the capture, but the majority of TMSIs were either a single occurrence, or were used for a short time.
\begin{table}[htbp]
\caption{Network Statistics}
\begin{center}
\resizebox{\columnwidth}{!}{%
\begin{tabular}{|c|c|c|c|}
\hline
\textbf{}&\multicolumn{3}{|c|}{\textbf{Network Operators}} \\
\cline{2-4} 
\textbf{Metrics} & \textbf{\textit{Operator 1}}& \textbf{\textit{Operator 2}}& \textbf{\textit{Operator 3}} \\
\hline
Total Pages& 586701 & 280795 & 156311\\
Unique TMSIs& 31654 & 36544 & 49076\\
Longest active TMSI in minutes& 361.25 & 361.04 & 288.15\\
\hline
\end{tabular}
}
\label{tab1}
\end{center}
\end{table}

Figure 5 shows the histograms of the lifespans of the observed TMSIs. Most TMSIs are very short lived, whereas some are observed for the entire duration of 6 hours. Our measurements were taken at a single location and we had no control of the UEs in the area. Due to the mobility of users, it is likely that the average TMSI lifespan is longer than shown here.

Operator 1 has a significant number of long-lasting TMSIs. This implies that many UEs attached to this cell did not hand-off connectivity during our experiements. This is displayed in Fig. 5 which shows a higher number of TMSIs at the maximum observed time for operator 1 when compared to the other two operators. 

\begin{figure}
    \centering
    \subfigure[Operator 1]
    {
        \includegraphics[width=3.0in]{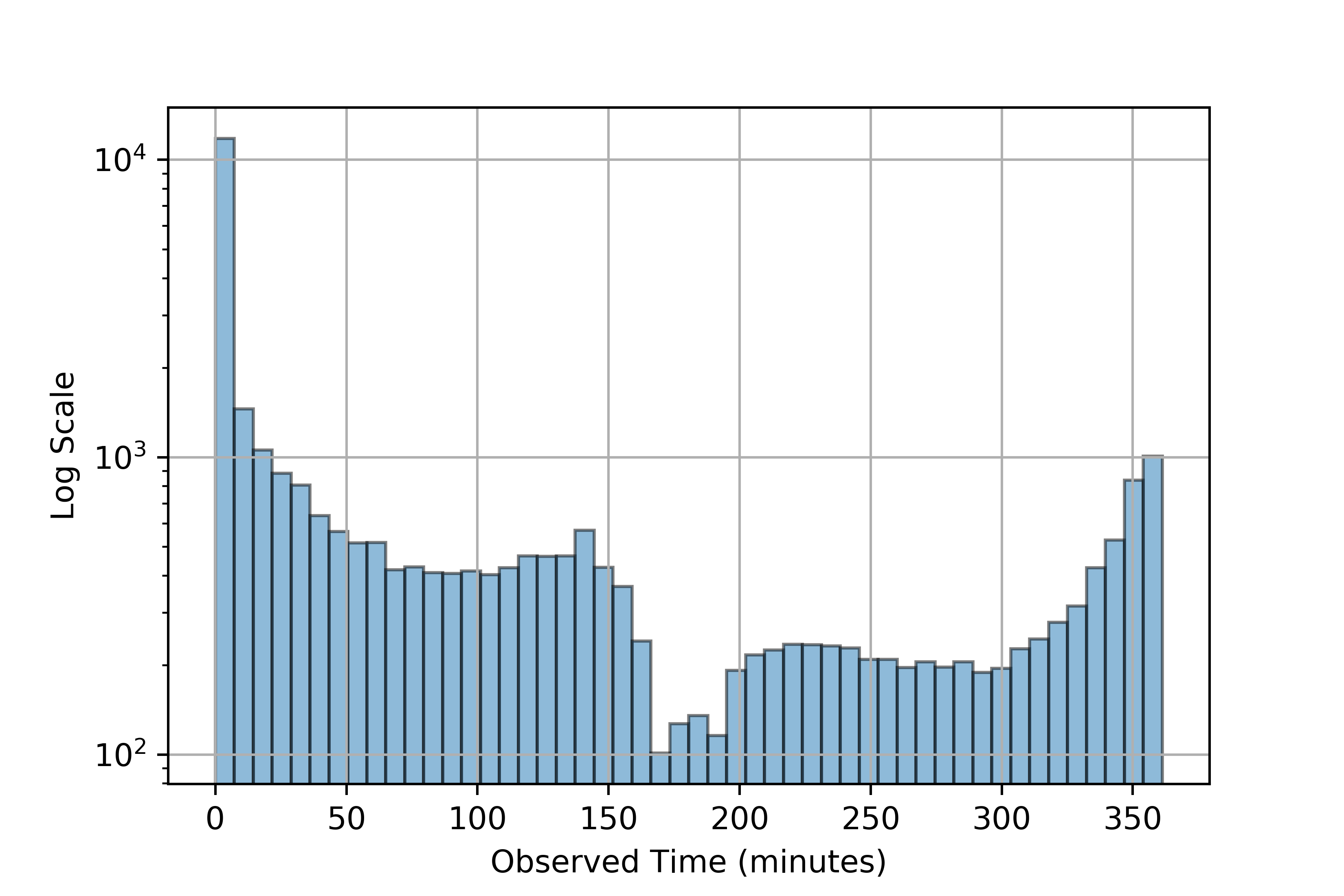}
        \label{fig:first_sub}
    }
    \\
    \subfigure[Operator 2]
    {
        \includegraphics[width=3.0in]{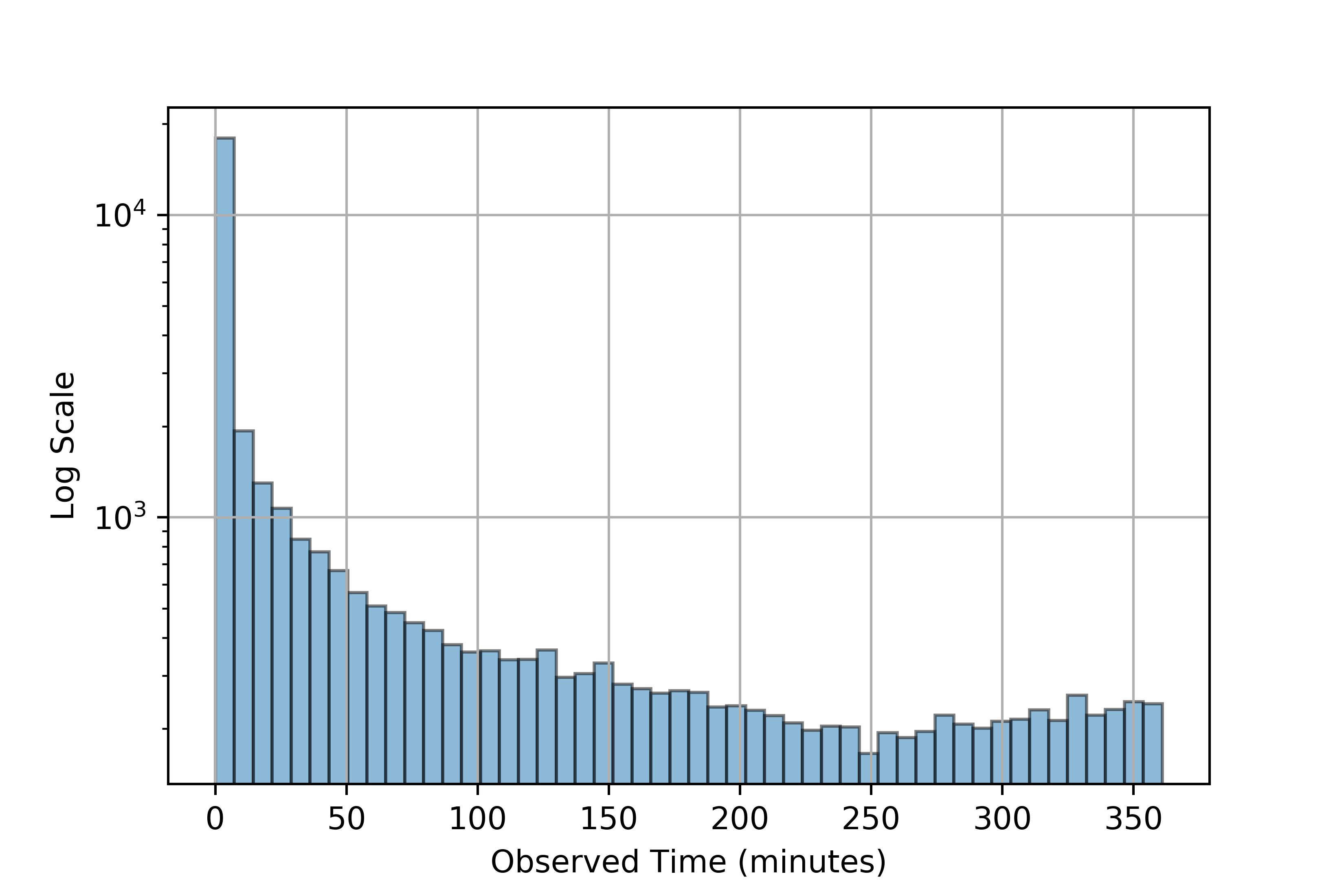}
        \label{fig:second_sub}
    }
    \subfigure[Operator 3]
    {
        \includegraphics[width=3.0in]{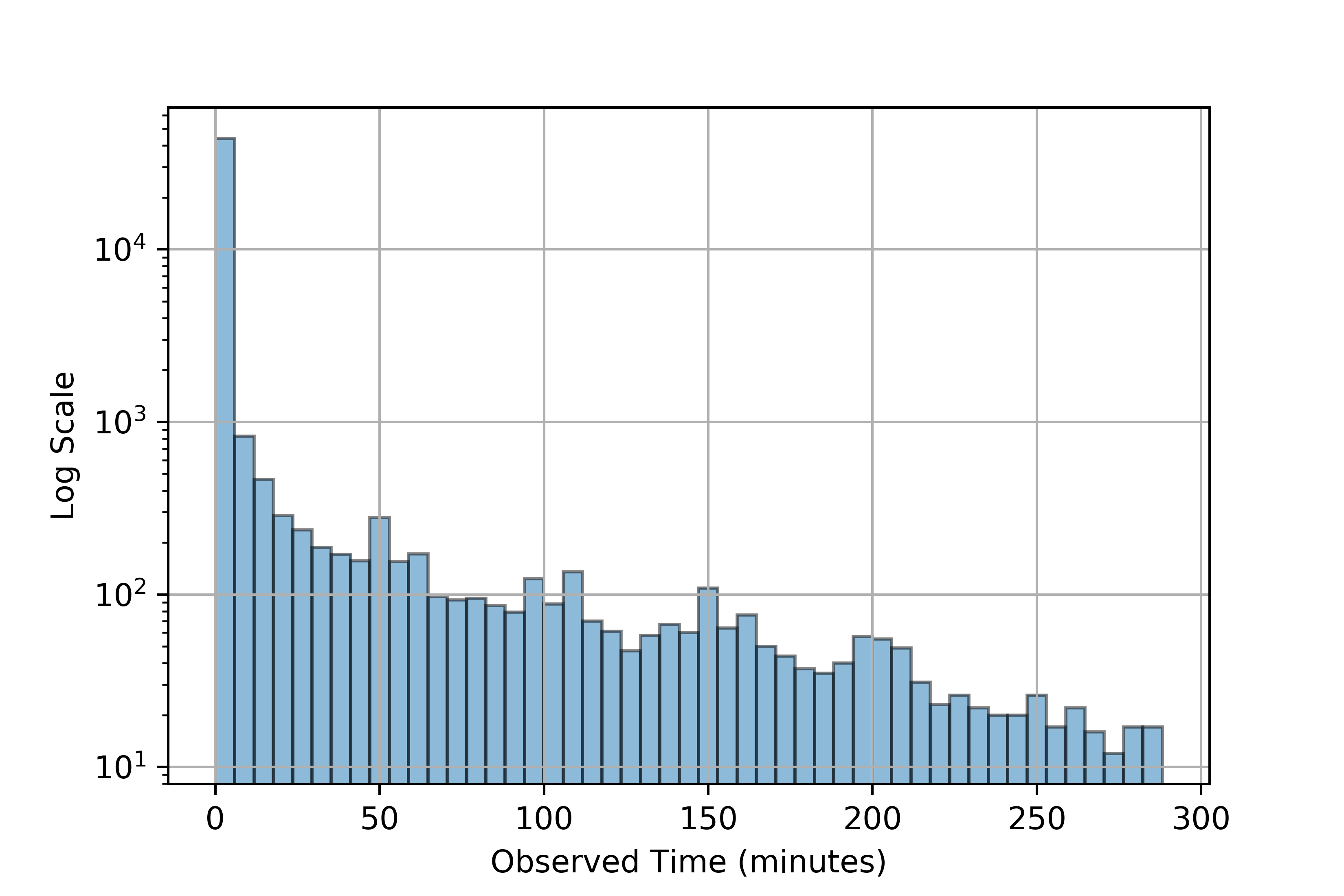}
        \label{fig:third_sub}
    }
    \caption{Time distribution of observed Paging messages.}
    \label{fig5}
\end{figure}

\subsection{Long-Term Persistence}

Next, we examine the persistence of TMSIs across different days. To test this, we use two commercial UEs and CSAI to monitor its operating band.  We initiated communication to the UE in attempts to capture the paging messages. This was accomplished by sending numerous messages from one UE to the other with ample time between messages so that the RRC connection is released due to UE inactivity between messages. The next day we repeated a similar communication pattern to generate more paging messages to our UE. 

We review the log files to see if our TMSI from day one persisted on the following day.  While the test was limited in scope, we did not observe any of the TMSIs from day one in the day two capture. This is a particularly encouraging result, as it implies that this network operator rotates TMSIs at least every twenty-two hours, which was the time window between our test captures. 

\section{Security Implications}
\subsection{Potential Attacks}
CSAI takes advantage of the inherent nature of pre-authentication and broadcast signaling in LTE.  While paging messages do not inherently contain sensitive information, it is possible to map a TMSI to a RNTI if you monitor subsequent RRC connection setup requests. Once a mapping is obtained, an attack as described in \cite{rupprecht2019breaking} could allow for statistical traffic analysis even though the contents of the NAS messages are encrypted and reflects a privacy concern.

Lichtman et al. outline attacks in \cite{lichtman2016lte} that discuss jamming of the LTE signals.  Once the MIB/SIBs are decoded, it is possible to target jamming efforts towards a specific eNB.  Combined with the aforementioned TMSI to RNTI mapping, it would be possible to extend the attack and jam one or several UE's data and control plane traffic. 
	
Most network operators will page a UE using a TMSI; however, 3GPP standards allow eNBs to page a UE using its IMSI in cases where a UE does not respond to three subsequent paging attempts using a TMSI.  This presents a significant security issue as many follow-on attacks are capable once a UE’s IMSI is known and include down bidding attacks or man in the middle style interceptions as demonstrated in \cite{shaik2015practical} and \cite{jover2016lte}. Our instrument enables research on security analysis and system hardening. Researchers will benefit from CSAI as they test modifications to 4G and 5G protocols to prevent the exploitation of preauthentication messages as surveyed in \cite{ferrag2018security}.

\subsection{Crashing a Software eNB}

Another interesting behavior that we observed in the course of developing CSAI was the potential for a straightforward Denial of Service attack against an SDR eNB through active RF attacks that mimic older Transmission Control Protocol (TCP) SYN Flood attacks. A SYN Flood attack exploits the inherit trust in TCP where a client floods a server with TCP SYN messages in the first stage of the TCP three way handshake. The server will allocate resources for the connection and reply to the client with a TCP SYN ACK message. Instead of completing the handshake with an ACK message, the client will disregard the server's SYN ACK and continue opening connections with the server.  This will eventually consume all resources on the server leading to a system crash or denying connectivity to legitimate clients. 

In modifying the \texttt{rrc.cc} file, if one sends an RRC Connection Request to an eNB and immediately calls the RRC Connection Release process, the UE will not respond to the eNBs request for RRC Connection Setup.  The eNB will allocate resources for the UE in expectance that the UE will reply with an RRC Connection Setup Complete; however, the UE has already begun the process of releasing the connection.  This leaves the base station in a half open state waiting for the UE to finish the RRC handshake.  Since the UE was instructed to release, after a short delay it will attempt to reconnect to the eNB further exhausting its resources.  An even faster way to perform this attack would be to have a fake UE enter a while loop that constantly requests and immediately releases RRC Connections.

In performing this attack against a SDR eNB, we were able to crash it with high success. In our investigation, the eNB crashed due to automated buffer overflow protections enabled by default when using the GNU C Compiler. An example of this crash is shown in Fig. 6. 

While more testing is required to determine the scope of this active attack, one potential mitigation may be similar to SYN cookies as detailed in \cite{Kurose:2016} where the eNB would only allocate resources for RRC Connections after the UE responds with the full Setup Complete message. In the case of performing this attack without the while loop, the eNB occasionally took minutes before it crashed.  It is possible to modify CSAI be modified to include a delay that ensures the eNB does not crash, but rather denies service to legitimate UEs that are connected or are trying to connect. This finding exemplifies the potential of fuzzing analysis against the cellular network infrastructure. In the case of the analysis presented in this manuscript, a software radio-based UE successfully crashed an open-source LTE network. However, the same could occur against a commercial eNB. We are currently investigating this further and impact on commercial femtocells and their protection against the attacks demonstrated in \cite{kim_ltefuzz_sp19}.

\begin{figure}[htbp]
\centerline{\includegraphics[width=\columnwidth]{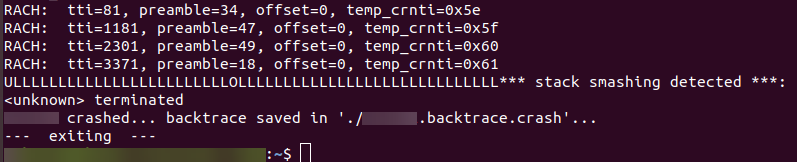}}
\caption{Crash of a software eNB}
\label{fig6}
\end{figure}

\section{Conclusion}
This paper describes how a SDR LTE implementation of a cellular RAN can be repurposed for analyzing the security of the wireless network. Our example is for 4G LTE, but similar principles can be applied to other cellular communications protocols. Using method two described in Section IV, SIBs and paging messages are able to be passively recorded from test or commercial LTE networks; we also discovered another modification to allow for DoS attacks against eNBs.  When 5G specifications are frozen and NR begins to deploy, this tool will be extended to capture those messages as well because the signaling in NR is similar to that of LTE/LTE-A.

We will release our code so that the community can utilize this instrument for ongoing investigations on RAN security. In continuing research, we are using CSAI as we investigate practical attacks and remediations for UE and eNB implementations. These include base station/small-cell fuzzing, location leakage, and UE denial of service attacks and their countermeasures.

\nocite{*}

\balance
\bibliographystyle{IEEEtran}
\bibliography{bib.bib}

% that's all folks
\end{document}